 \documentclass[preprint]{aastex}

\slugcomment{submitted to \aj}

\shorttitle{C stars in NGC 6822}
\shortauthors{Letarte \& al.}

\begin{document}


\title{The extent of NGC 6822 revealed by its C star population}

\author{Bruno Letarte\altaffilmark{1} and Serge Demers\altaffilmark{1}}
\affil{D\'epartement de Physique Universit\'e de Montr\'eal, Montreal, Qc
H3C 3J7, Canada}
\email{letarte@astro.umontreal.ca; demers@astro.umontreal.ca}

\author{Paolo Battinelli\altaffilmark{1}}
\affil{Osservatorio Astronomico di Roma, viale del Parco Mellini 84, I-00136 Roma, Italy}
\email{battinel@oarhp1.rm.astro.it}
\and
\author{W. E. Kunkel}
\affil{Observatories of the Carnegie Institution of Washington, Casilla 601,
La Serena, Chile}
\email{kunkel@jeito.lco.cl}
\altaffiltext{1}{Visiting Observers, Canada-France-Hawaii Telescope, 
operated by the National Research Council of Canada, le Centre National
de la Recherche de France and the University of Hawaii.}

\begin{abstract}
Using the CFH12K camera, 
we apply the four band photometric technique to identify 904 carbon stars
in an area $28'\times 42'$
centered on NGC 6822. A few  C stars, outside of this area, were also
discovered with the Las Campanas Swope Telescope. The NGC 6822 C star 
population has $<I>$ = 19.26 leading to 
a $<M_I>$ = --4.70, a value essentially identical to the mean 
magnitude obtained for the C stars in IC 1613. 
Contrary to stars highlighting the optical image of NGC 6822,
C stars are seen at large radial distances and trace a huge slightly
elliptical halo which does not coincide with the 
huge  HI cloud surrounding NGC 6822. The previously 
unknown stellar component of NGC 6822 has a exponential 
scale length of $3.0'\pm 0.1'$ and can be traced to five scale lengths.
The C/M ratio of NGC 6822 is evaluated to be 1.0 $\pm 0.2$.

\end{abstract}



\section{INTRODUCTION}
NGC 6822 is a barred dwarf galaxy of type IrIV-V  (van den Bergh 1968), 
a classification similar to the Small Magellanic Cloud. 
Hubble (1925) discovered Cepheids in NGC 6822 and could conclude that it is
a relatively nearby galaxy. 
NGC 6822 has recently been the subject of detailed multi-color investigation by 
Gallart, Aparicio \& V\'\i lchez (1996).  We
adopt, for NGC 6822, their determined color excess E(B--V) = 0.24 $\pm 0.03$
and true modulus $(m-M)_o = 23.49\pm 0.08$. The abundance of two supergiants
of NGC 6822 has recently been determined, from high resolution spectroscopy,
by Venn et al. (2001). We adopt their estimate, [Fe/H] = --0.49, for the
metallicity of NGC 6822.

NGC 6822 is nearly unique among the
Local Group galaxies by having a huge hydrogen envelope, several times
bigger that its optical core (Robert 1972, Gottesman \& Weliachew 1977, 
de Blok \& Walter 2000). IC 1613 (Lake \& Skillman 1989), IC 10 the
highly reddened galaxy (Huchtmeier 1979) and to a lesser extent, NGC 3109
(Barnes \& de Blok 2001) have also a similar HI envelope.
The global structure of NGC 6822 was first surveyed by Hodge (1977) 
and in more detailed by Hodge et al. (1991) who could trace
the galaxy to 10$'$ from its center. Hodge adopted the point of view that
the outer structure of NGC 6822 is circular rather than following the
shape of the bar. Because  NGC 6822 is at a rather low Galactic 
latitude, b = --18.4$^\circ$, the low density periphery of the galaxy
is masked by the foreground stars. The exact size of NGC 6822 is rather
difficult to establish simply from star counts.
The NASA/IPAC Extragalactic Database (NED) quotes a dimension of
$15.5'\times 13.5'$ which is much smaller than the HI envelope mentioned above
but is quite in line with Hodge's study.

All photometric surveys of NGC 6822,
with the exception of the one by Hutchings, Cavanagh \& Bianchi (1999) have
targeted the obvious central part of the galaxy. Hutchings et al. (1999) have
obtained HST photometry in two small regions located $\sim$ 2 kpc from the
optical center of NGC 6822. They detected the presence of main sequence
stars extending to 2 M$_\odot$ in the western field and suggested that this
population might be the result of a tidal event. The presence of an hydrogen
tidal arm was proposed by the HI map of de Blok \& Walter (2000).

Cook, Aaronson \& Norris (1986), who pioneered the photometric technique
we are now using, were the first to target NGC 6822 in search of C stars. 
They observed two small ($1.5'\times 2.5'$) fields located along the central
bar on opposite sides, a few arcminutes from its geometrical center.
Some fifty C stars were identified. Follow-up spectroscopy
 confirmed the nature
of the C stars (Aaronson et al. 1984) and lead to the discovery of the first
S star in NGC 6822 (Aaronson, Mould \& Cook 1985). That investigation 
observed barely 3\% of the area of NGC 6822, if we assume the dimensions
given by NED.

Battinelli \& Demers (2000) have obtained a preliminary relation linking
the number of C stars in a galaxy with its integrated absolute visual 
magnitude. This relation predicts that galaxies with $M_V < -15$ would
have more than a few hundred C stars. NGC 6822 with
$M_V = -15.2$ (Mateo 1998) should have at least 500 C stars. Such large
number would provide sufficient statistics to investigate the extent
of NGC 6822.
A survey of the periphery is particularly interesting for a galaxy such as
NGC 6822. Indeed, tidal interaction has been proposed (de Blok \& Walter
2000) to explain the asymmetric shape of the 
 extended hydrogen disc. It is reasonable to expect that the spatial 
distribution (and kinematics) of stars as old
as $\sim 3$ Gyr could also have been influenced by the presumed tidal
interaction. On the other hand, the huge HI envelope, currently visible,
implies that, if a gravitational interaction took place, it must not have
been strong enough to drastically disturb the gas. In this respect, C stars
have a distinctive advantage, over deep multi color surveys, 
because the foreground contribution is zero!
  
On a more global scale, 
in order to compare the C star population of different galaxies,
 one must be 
careful to compare homogeneous samples. That is, not to collect from the
literature lists of C stars identified spectroscopically of from 
different photometric criteria. For example, the $B_j -R$ criterion
used by Totten \& Irwin (1998), Whitelock, Irwin \& Catchpole (1996) or 
Kunkel, Irwin \& Demers (1997), to select C star candidates for spectroscopy,
is not at all equivalent to the 
(CN -- TiO) criterion adopted by Brewer, Richer \& Crabtree (1995) and
our series of investigations. This is why this paper follows strictly
the photometric criteria previously used. In this way, the same C star
population will be revealed in a number of dwarf galaxies of 
different properties. 

\section{OBSERVATIONS AND DATA REDUCTION}
\subsection{The Las Campanas data}
The NGC 6822 survey employs two set of observations obtained by rather
different telescopes and with different procedures. The Swope telescope, on Las
Campanas, was used in October 1999 to secure images of NGC 6822 in Kron-Cousins
R$_{KC}$, CN (810 nm)
and TiO (770 nm) filters. The goal of this brief survey was to identify
C stars, from the color-color diagram, for follow-up spectroscopy. 
 Thus, it was deemed unnecessary to 
acquire I images. A pseudo i magnitude was calculated by adding the CN 
and TiO magnitudes. The narrow interference filters, used here, 
 have a FWHM of 30 nm and are
the ones used in our previous surveys (Albert et al. 2000; Battinelli \&
Demers 2000; Demers \& Battinelli 2002). 
The $2048\times 2048$ SITe\#1 CCD  yields a
field of view of $23.7'\times 23.7'$. Two slightly overlapping fields were
observed for a total of five hours. Dome flats were obtained through each filters.
Calibration to the standard R system was done with Landolt's (1992) equatorial 
standards observed during the course of the nights. This was done simply to
set the magnitude zero point, not as an absolute color calibration.
No calibration is done for the CN and TiO filters because we are interested
only in the CN -- TiO color index and their exposure
times are always equal.

After the standard prereduction of trimming, bias subtraction, and sky
flat-fielding,
the photometric reductions were done by fitting model point-spread
functions (PSFs) using DAOPHOT/ALLSTAR/ALLFRAME series of programs (Stetson
1987, 1994) in the following way: we combine, using MONTAGE2, all the images
of the target irrespective of the filter to produce a deep image devoid of
cosmic rays.
ALLSTAR was then used on this deep image to derive a list
of stellar images and produce a second image where the stars, found in the
first pass, are removed. This subtracted image is also processed through
ALLSTAR to find faint stars missed in the first pass. The second list of
stars is added to the first one. The final list is then used for the analysis
of the individual frames using ALLFRAME. This program fits model PSFs to
stellar objects in all the frames simultaneously.
The journal of these observations is presented in Table 1.
\placetable{tab-1}
\subsection{The CFHT data}
NGC 6822 was also observed with the CFH12K at the beginning of one night in
September 2000. The camera consists in a 12K$\times$8K pixel mosaic 
covering a field
of $42'\times 28'$. Each pixel corresponding to 0.206 arcsec. 
Images were obtained through Mould I,R filters and CN,TiO filters.
The CFHT narrow band filters have, however, about half the bandwidth
of the Las Campanas ones, thus requiring relatively longer exposure times.
The journal of these  observations is also given in Table 1.

Preanalysis of our data was done with the FITS Larges Images Processing
Software (FLIPS) package at the CFHT Headquarters.  It is a program specially
designed to handle mosaic data.  
FLIPS corrects for the bias, dark and flat, by
averaging the good exposures and rejecting the overexposed and otherwise
incorrect frames, for each of the 12 CCD fields of the mosaic.  
It also takes into account the mask of bad pixels.
A nice feature of FLIPS, compared to standard pre-reduction packages, is
that it normalizes the background sky values to the most sensitive CCD of
the mosaic.  This results in 12 fields that are comparable and on the same
magnitude scale.
 
There were a fringe patterns in a few CCD I frames, but no correction for the
fringing was done.  FLIPS allows us to correct for the fringing effect but
``no correction at all'' was deemed  the best option.  
Trying to correct for
that cosmetic problem often increases the noise on the frame.  On a small
scale, fringing was fairly constant and in the end, our analysis packages
had no difficulty accounting for that anomaly.
 
Data reduction was done with DAOPHOT/ALLSTAR, in a way similar to the
reduction of the 
Las Campanas data. 
\subsubsection{The CFH12K calibration}
The CFHT observations were unfortunately obtained under non photometric
conditions, even though the seeing was less than one arc second. We therefore
had to used local stars with published magnitudes and colors to calibrate
our photometry. This, however,
presented a challenge because very few publications present R--I colors. We were
fortunate to obtain the NGC 6822 photometry of Gallart et al. (1996) which 
satisfies our needs entirely. 

Using 166 stars, with colors in the range $0.0 < (R-I) < 2.0$, we obtained
the following transformation equations:
$$ R = r + 7,643\pm 0.022 + 0.0040\pm 0.0196(R-I)$$
$$ I = i + 7.200\pm 0.009 + 0.0126\pm 0.0128(R-I)$$
The r and i magnitudes are the instrumental magnitudes 
calibrated to Gallart et
al. (1996) photometry.
We thus conclude that there is no color term and that the Mould filters 
provide an excellent match to the Kron-Cousins magnitudes.

The non-photometric conditions oblige us to set a zero point to the CN--TiO
colors, even though exposures in each filter were of the same length. We do
so by following Brewer et al. (1996) and setting the mean of the (CN--TiO)
= 0.00
for all stars with (R--I)$_o < 0.45$ since hot stars are expected to have
featureless spectra in the CN or TiO regions. Brewer et al. (1996) used
V--I colors, the R--I limit adopted here takes into account the relationship
between the two indices. Photometry of each of the twelve
CCDs was thus calibrated using several hundred stars in each CCD. Stars
in the color range $0.2 < (R - I) < 0.65$ were selected for this zero point
adjustment, taking into account the E(R--I) = 0.20.

\section{RESULTS}
\subsection{The color-magnitude diagram}
The global color magnitude diagram, (CFH12K data)
 containing some 65000 stars, is shown in
Figure 1. Only stars with small photometric errors are plotted here. This
explains why the fainter limits  abruptly appears at I$\approx$ 21.5, 
{\it two magnitudes below the red giant tip}. 
We restrict our analysis to stars for
which the square root of the quadratic  sum of the errors on R--I and 
CN--TiO is less than 0.100. This limit is somewhat arbitrary but is justified
because
 the inclusion of numerous fainter stars pollutes enormously the color-color
diagram without increasing the number of carbon stars. 
\placefigure{fig1}
The vertical ridge, at $(R - I)\approx 0.55$ is an interesting feature usually
present in deep CMD fields toward low Galactic latitudes. The blue side of this
ridge corresponds to the main sequence turnoff of field G dwarfs. If we adopt
$(R - I)_\circ$ $\approx$ 0.35 for G5 dwarfs (Cox 2000), then the E(R -- I) 
toward NGC 6822 would be $\approx$ 0.20, a value identical to the one proposed
by Gallart et al. (1996)

We can see right away, without the analysis of the space distribution of
the C stars, that
NGC 6822 extends much further than its optical image makes us believe. Figure 2
displays the CMD of the periphery of our $42'\times28'$ field. 
Only stars farther
than $17.1'$ ($\sim 2.5$ kpc) from the center of the field are plotted. 
A weak giant branch is 
present. The true extent of NGC 6822 is however quite difficult to assess from
its CMD because of the heavy foreground contribution by Galactic stars. 
As we shall see, C stars are valuable
in this respect because none are in the foreground!
\placefigure{fig2}
\subsection{The color-color diagram}
The color-color diagram of the whole CFH12K field of NGC 6822 is presented 
in Figure 3.
904 stars satisfy our criteria and are called C stars. We define, for the purpose
of comparison from galaxies to galaxies, C stars as stars with $(R -I)_o >
0.90$ and, in the case of narrow CN and TiO filters, stars with $(\rm{CN} - 
\rm{TiO}) > 0.30$. This $(R-I)_o$ limit corresponds to spectral type M0,
according to Bessel (1991).
Battinelli \& Demers (2000) and Aaronson et al. (1985) 
showed that (CN -- TiO) is very little affected by
reddening. This definition is obviously restrictive, one can easily 
see that on,
Fig. 3, there are bluer C stars extending to the lower left of the C star 
We discuss those bluer stars in section 4.2
The C stars identified, along with their J2000.0 equatorial
coordinates, magnitudes and colors,  are listed in Table 2, (only the
first few stars are given in the paper version). Star 
c236  is the
one spectroscopically confirmed by Aaronson et al. (1984).
\placefigure{fig3}
\placetable{tab-2}
\subsection {Comparison of the Las Campanas and CFHT results}
The Las Campanas photometry (LC) yielded 688 C star candidates. 
A coordinate match with
the CFHT candidates shows that only 542
 are common to both sets. A comparison of
the apparent magnitude of stars common to both sets, displayed in Figure 4, shows
that a substantial number of faint C stars are missing from the LC data set. The
number of C star candidates, in the LC list,  could have been increased by lowering the 
acceptance criteria based on the photometric errors. We believe, however,
that the use of the combined CN and TiO magnitudes to produce
a pseudo I magnitude has lowered  the photometric quality of I 
magnitudes compared to R magnitudes of similar brightness. 
This approach has obviously raised the acceptable magnitude of the LC data
set but it has allowed us to identify the C star population of NGC 6822, for 
follow-up spectroscopy, more than one year before we obtain 
the CFHT observations
confirming  the existence of a huge C star population.
\placefigure{fig4}
About 100 stars were identified as C stars in the LC data set but
were not retained in the CFHT list. A search and match in the whole
CFH12K database revealed that nearly all them can be matched to stars. Some
of them are just a little bit too blue to be called C stars or have a 
(CN--TiO) index a little too small. The majority, however, can match in
position, fainter stars of various colors. This suggests that the LC
stars are actually blends. The pixel scale of the Swope telescope CCD 
is not
as good as the one of the CFHT. The LC fields extend further west than the 
CFH12K mosaic. Three possible C stars outside of the CFH12K field
 are listed in Table 3. These 
candidates need to be confirmed by more accurate photometry. They are
located on the edge of the Swope Telescope field and their photometric errors 
are, in some cases, near the limit of rejection.
\placetable{tab-3}

\section{DISCUSSION}
The mean properties of the carbon star population of NGC 6822 are nearly
identical to the one of IC 1613, a dwarf irregular galaxy of lower mass
but with a substantial C star population.
The mean magnitude and color of the 904 C stars are: $<I>$ = 19.257 and
$<(R - I)>$ = 1.368, corresponding to $<M_I>$ $ = -4.70$ and $<(R-I)_o>$
$ = 1.17$. Albert et al. (2000) quote $<M_I>$  = --4.69 and 
$<(R-I)_o>$ = 1.18 for the 195 C stars
in IC 1613. The 
abundance of IC 1613 was estimated, from the color of the tip of
the giant branch, by Freedman (1988) to be [Fe/H] = --1.3, a value lower than the
adopted metallicity 
for NGC 6822. This comparison suggests that the metallicity
may have little effect on the mean M$_{I}$ of a C star population. 
\subsection{The Bolometric magnitude distribution of C stars}
Costa \& Frogel (1996) were able to determine the bolometric magnitude
of C stars in the Large Magellanic Cloud (LMC) from their R, I photometry.
Since the newly established metallicity of  NGC 6822 (Venn et al. 2001) is
quite close to the  metallicity of the LMC, we may 
assume that their equation also applies to the NGC 6822 C stars.
Figure 5 displays the bolometric magnitudes of C stars, listed in Table 2,
as a function of their (R -- I)$_o$. Our distribution, when compared to
the one of Costa \& Frogel (1996) for the LMC, is truncated on the
blue side due to our adoption criterion. Figure 5 does show, however, that
NGC 6822 does  contain one C star brighter than M$_{bol} = -6.4$, 
the limit according to 
 models of Boothroyd, Sackmann \& Ahern (1993). That star is c043, a bright
(in I) star located near the center of NGC 6822.
\placefigure{fig5}
\subsection{Bluer ``C stars''}
Demers \& Battinelli (2002) have recently shown that several 
spectroscopically identified C stars in the Leo I dwarf spheroidal
galaxy are bluer than our adopted (R--I)$_o$ limit. There could then 
certainly be
a number of bluer C stars in NGC 6822. Indeed,
one can easily see on the color-color diagram (Fig. 3)
that there is a natural blue extension of the ``C star branch''.
In order  to compare the photometric properties of these
stars, with the previously defined C stars, we select, on the color-color
diagram, stars in the following box: 0.8 $<$ (R--I) $<$ 1.1 and 
(CN--TiO) $>$ 0.25. 341 stars are in this region of the color-color
diagram. As expected, these bluer stars are located on the nearly vertical
part of the red giant branch. Figure 6 presents a close-up of the CMD and 
histograms comparing the magnitude distributions of the two groups of stars. 
Contrary to redder C stars, which have a narrow range of magnitude, 
the bluer stars show a large magnitude spread.  
\placefigure{fig6}
\subsection{The known S star}
Aaronson et al. (1985) spectroscopically confirmed the presence of at least
one S star in NGC 6822. This star is in our database but its color
indices are such that our color-color diagram cannot really distinguish it
from M stars. Its magnitude and colors are: I = 18.346; (R--I) = 1.340;
(CN--TiO) = 0.056. The bolometric magnitude of this S star, using the
same relation than for C stars, is M$_{bol}$ = --5.90. Brewer et al. (1996)
found a single S star among their AGB stars with a surprisingly high
bolometric magnitude, --6.2. 

In our coordinate system, the J2000.0 coordinates this S star are
some 10 arcsec from the ones quoted by Aaronson et al. (1985). They
are: $\alpha = 19^h 45^m01.3^s$, $\delta = -14^\circ 49' 32.1''$. There could
be other S stars in NGC 6822, spectroscopy would be needed to confirm the
nature of the stars lying in between the C and M regions on the color-color
diagram.
\placefigure{fig7}
\subsection{The spatial distribution of C stars and red giant stars}
We display, in Figure 7, the 904 C stars of NGC 6822 
over a $42'\times 28'$ ESO Digital Sky Survey (DSS) Image centered on NGC 6822. 
C stars reveal that 
the stellar population of 
NGC 6822 extends over a much larger volume than its optical image 
would make us believe.  Furthermore, C stars are not restricted to the 
HI disk, nicely mapped by de Blok \& Walter (2000), but are found in 
a spheroidal halo. 
Radial velocities will help establishing if C stars follow or not
the HI rotation. 
\placefigure{fig8}
The surface density profile of the C stars, shown in Figure 8, is well fitted
by a power law with a scale length of 3.0 $\pm 0.1$ arcmin, 
corresponding to 436 pc at the distance of NGC 6822. We assume circular
symmetry. The exponential profile can be followed to five scale lengths.
This scale length is to be compared with $112''\pm 50''$ given by Hodge et al.  (1991). 

The fact that we find intermediate age stars in the outer halo of a dwarf
galaxy is surprising since the halos so far known around dwarf galaxies
(Lee 1993, Minniti \& Zijlstra 1997, Minniti, Zijlstra, \& Alonso 1999,
Aparicio \& Tikhonov 2000, Aparicio, Tikhonov \& Karachentsev 2000) 
consist of $\sim 10$ Gyr
stars. We must stress, however, that a weak intermediate age population
will be missed by a CMD approach since there is no way to distinguish
intermediate and old age red giants. 

Photometric properties of C stars, close to
the center and in the periphery of NGC 6822, appear 
to be essentially the same. We divide the data into four radial bins to calculate
mean magnitudes and mean colors. We find dispersions of $\pm 0.01$ for $<I>$ 
and $\pm 0.02$ for $<$(R--I)$>$. 

Since Fig. 1 shows a well defined upper giant branch, one can
select a representative sample of NGC 6822 giant stars from their position 
on this CMD. This would represent a mixture of old and intermediate age stars. 
We select, from the
CMD of the whole field, giant stars in a narrow parallelogram in the 
magnitude range 19.5 $<$ I $<$ 21.3 and with the appropriate R--I intervals,
slightly variable with magnitude to take into account the sloping giant
branch.
More than 18000 stars are in this box, including a number of foreground
stars.  The smoothed surface density map of the giants is presented in Figure 9. 
No correction for the foreground contribution
has been done here. An obvious asymmetry is seen, along the E-W axis,
 in the inner isodensity contours. 
The outermost contour, corresponding
to a surface density of 13.5 stars per arcmin$^2$, is roughly
elliptical with an ellipticity of e $\approx$ 0.1 and a position angle
of PA $\approx 60^\circ$. This is quite different from the orientation
of the HI disc (PA $\approx 125^\circ$) 
and not aligned with the optical bar which is, according to Hodge (1977),
at PA = $10^\circ \pm 3^\circ$. The outermost contour has a major
axis of $\sim 23'$ or 3.3 kpc at the distance of NGC 6822. The estimated foreground
contribution to giant counts, detailed below, amount to less than 25\% of the 
counts in the outermost contour.

\placefigure{fig9}

Since it appears that the ellipticity of the halo of NGC 6822 is small
in the outer parts and not evident for the inner contours, we decided
to neglect it and compute radial counts in circular annuli, like we did
for the C stars.  However, for red giants, the foreground pollution is
not negligible and must be taken into account in the outer parts.
We estimate it from star counts in two strips  on the eastern 
and western sides of the CFH12K field. We obtain: 3.5 $\pm 0.2$ stars per arcmin$^2$. 
The scale length, obtained by least-square
fit to the middle points of Figure 8, is $3.3' \pm 0.2$ a value similar
to the one obtain from C stars. We can thus conclude that the old and
the intermediate age halos have the same size and show no pronounced
asymmetry like the hydrogen cloud surrounding NGC 6822.

\subsection{The C/M ratio}
The size of NGC 6822, relative to our mosaic, is such that it is quite
difficult to evaluate the foreground contribution to the M stars seen
in the color-color diagram. Furthermore, the fact that NGC 6822 is at
a relatively low Galactic latitude makes the problem even worst.
We evaluated the foreground by counting M stars (with I $<$ 21.0)  
in two 1000 pixel wide strips on the
eastern and western extremities of the field. 
Their numbers were 746 and 717 respectively.
We have of course to assume that
the NGC 6822 contribution is negligible in these areas. This is not 
strictly true since we see  a few C stars up to the edge of the field. 
The total number
of M stars (I $<$ 21.0) in the whole field is 9930. 
From the two counts above, we
estimate that there are 8989 $\pm 184$ foreground M stars.
The number of M stars within 
NGC 6822 is thus sensibly similar to the number of C stars and equal 
to 941 $\pm 184$, taking into account this uncertainty,
we obtain a global C/M = $1.0\pm 0.2$. The huge number of foreground 
M stars, compared to the NGC 6822 M stars, makes hopeless
any attempt to investigate
a radial dependence of C/M. One would need a more accurate
evaluation of the foreground contribution from star counts 
in regions outside of our field.  The M stars counted are those
with spectral type M0 or later that have (R--I)$_o > 0.90$.

\acknowledgments
We are grateful to Carme Gallart to have provided her NGC 6822 photometry
to enable us to calibrate the CFH12K data. 
This project is supported financially, in part, by the Natural Sciences and
Engineering Research Council of Canada (S. D.).


\clearpage
%
%



\figcaption[letart.fig1.ps]{Color-magnitude diagram of the whole CFH12K field,
centered on NGC 6822.
\label{fig1}}

\figcaption[letart.fig2.ps]{Color-magnitude diagram of the periphery of the field, farther
than 17.1$'$ from the center of NGC 6822. We can still see a weak giant
branch.
\label{fig2}}

\figcaption[letart.fig3.ps]{Color-color diagram of the $42'\times 28'$ field,
hundreds of C stars are present. Nearly 65000 stars are plotted.
\label{fig3}}

\figcaption[letart.fig4.ps]{I magnitude distribution of the C stars listed in 
Table 2. The shaded histogram corresponds to the 542 C stars common to 
both Las Campanas and CFHT data. It reveals the incompleteness of 
the LC data at faint magnitudes.
\label{fig4}}

\figcaption[letart.fig5.ps]{The bolometric magnitude of C stars as a function
of their intrinsic colors.
\label{fig5}}

\figcaption[letart.fig6.ps]{A close-up of the CMD is shown in the upper panel.
Dots are the 904 C stars while crosses are the bluer stars on the C stars ``branch''.
The magnitude distributions of the two groups of stars are compared in the lower
panel.\label{fig6}}
\figcaption[letart.fig7.ps] {C stars are plotted over the image of NGC 6822
from the DSS. The size of the field corresponds to the CFH12K. North is
on top and East is on the left. Three C stars, listed in Table 3, are
just outside the western limit.
\label{fig7}}

\figcaption[letart.fig8.ps]{Surface density of profiles: the 904 C stars are represented
by open circles while the red giants are shown as solid dots. Scale lengths are 
essentially the same.
\label {fig8}}
\figcaption[letart.fig9.ps]{Smoothed isodensity contours of the red giants
brighter than I = 21.3. No correction has been done for the foreground stars.
The orientation and size of the field are like the ones of Figure 7.
\label{fig9}}


\clearpage


\begin{thebibliography}{}
\bibitem[Aaronson et al. 1985]{aar85} Aaronson, M., Mould, J., \& Cook, K. H.
1985, \apj, 291, L41
\bibitem[Aaronson et al. 1984]{aar83} Aaronson, M., Da Costa, G. S., Hatigan, P,
Mould, J. R., Norris, J., \& Stockman, H. S. 1984, \apj, 277, L11
\bibitem[Albert et al. 2000]{alb00} Albert, L., Demers, S., \& Kunkel, W. E.
2000, \aj, 119, 2780 (Paper I)
\bibitem[Aparicio \& Tikhonov 2000]{apa00} Aparicio, A., \& 
Tikhonov, N. 2000, \aj, 119, 2183
\bibitem[Aparicio et al. 2000]{apa00} Aparicio, A., Tikhonov, N. \& 
Karachentsev, I. 2000, \aj, 119, 177
\bibitem[Barnes \& de Blok 2001]{bar01} Barnes, D. G. \& de Blok, W. J. G.
2001, \aj, 122, 825
\bibitem[Battinelli \& Demers 2000]{bat00} Battinelli, P. \& Demers, S. 
2000, \aj,
120, 1801 (Paper II)
\bibitem[Bessel 1991]{bes91} Bessel, M. S. 1991, \aaps, 88, 325
\bibitem[Boothroyd et al. 1993]{boo93} Boothroyd, A. I., Sackmann, I. J., \&
Ahern, S. C. 1993, \apj, 416, 762
\bibitem[Brewer et al. 1995]{bre95} Brewer, J. P., Richer, H. B., \&
Crabtree, D. R. 1995, \aj, 109, 2480
\bibitem[Brewer et al. 1996]{bre96} Brewer, J. P., Richer, H. B., \&
Crabtree, D. R. 1996, \aj, 112, 491
\bibitem[Cook et al. 1986]{coo86} Cook, K. H.,  Aaronson, M., \& Norris, J.
1986, \apj, 305, 634
\bibitem[Costa \& Frogel 1996]{cos96} Costa, E., \& Frogel, J. A. 1996, \aj, 112, 2607
\bibitem[Cox 2000]{cox00} Cox, A. N. 2000 Allen's Astrophysical Quantities,
Spinger-Verlag New York 2000, p392
\bibitem[de Blok \& Walter 2000]{deb00} de Blok, W. J. G. \& Walter, F. 
2000, \apj, 537, L95
\bibitem[Demers \& Battinelli 2001]{dem01} Demers, S. \& Battinelli, P. 2002,
\aj, 123, in press
\bibitem[Freedman 1988]{fre88} Freedman, W. L. 1988, \aj, 96, 1248
\bibitem[Gallart et al. 1996]{gal96} Gallart, C., Aparicio, A. \& V\'\i lchez,
J. M. 1996, \aj, 112, 1928
\bibitem[Gottesman \& Weliachew 1977]{got77} Gottesman, S. T. \& Weliachew, L. 1977, \aap,
61, 523
\bibitem[Hodge 1977]{hod77} Hodge, P. W. 1977, \apjs, 33, 69
\bibitem[Hodge et al. 1991]{hod91} Hodge, P. , Smith, T., Eskridge, P.,
MacGillivray, H., \& Beard, S. 1991, \apj, 379, 621
\bibitem[Hubble 1925]{hub25} Hubble, E. P. 1925, \apj, 62, 409
\bibitem[Hutchings et al. 1999]{hut99} Hutchings, J. B., Cavanagh, B,, \&
Bianchi, L. 1999, \pasp, 111, 559
\bibitem[Hutchmeier 1979]{hut99} Huchtmeier, W. K. 1979, \aap, 75, 170
\bibitem[Kunkel et al. 1997]{kun97} Kunkel, W. E., Irwin, M. J., \& 
Demers, S. 1997, \aaps, 122, 463
\bibitem[Lake \& Skillman 1989]{lak89} Lake, G. \& Skillman, E. D. 1989,
\aj, 98, 1274
\bibitem[Landolt 1992]{lan92} Landolt, A. U. 1992, \aj, 104, 340
\bibitem[Lee 1993]{lee93} Lee, M. G. 1993, \apj, 408, 409
\bibitem[Mateo 1998]{mat98} Mateo, M. 1998, \araa, 36, 435
\bibitem[Minniti \& Zijlstra 1997]{min97} Minniti, D. \& Zijlstra  1997, \aj, 114, 147
\bibitem[Minniti et al. 1999]{min99} Minniti, D., Zijlstra, A. A. \& Alonso, 
M. V. 1999, \aj, 117, 881
\bibitem[Robert 1972]{rob72} Robert, M. S. 1972, in {\it External Galaxies
and Quasi-stellar Objects} (IAU Symp. No. 44) ed. D. S. Evans, Reidel, 
Dordrech, p12.
\bibitem[Stetson 1987]{ste87}Stetson, P. B. 1987, \pasp, 99, 191
\bibitem[Stetson 1994]{ste94} Stetson, P. B. 1994, \pasp, 106, 250
\bibitem[Totten \& Irwin 1998]{tot98} Totten, E. J., \& Irwin, M. J. 1998,
\mnras, 294, 1
\bibitem[van den Bergh 1968]{vdb68} van den Bergh, S. 1968, Commun. David Dunlap Obs., 195
\bibitem[Venn et al. 2001]{ven01} Venn, K. A., Lennon, D. J., Kaufer, A.,
McCarthy, J. K., Przybilla, N., Kudritzki, R. P., Lemke, M., Skillman, E. D.,
\& Smartt, S. J. 2001, \apj, 547, 765
\bibitem[Whitelock et al. 1996]{whi96} Whitelock, P. A., Irwin, M. J., \&
Catchpole, R. M. 1996, New Astronomy, 1, 57
\end{thebibliography}
\end{document}